\documentclass[10pt,twocolumn,english]{article}
\usepackage[T1]{fontenc}
\usepackage[latin9]{inputenc}
\usepackage{babel}
\usepackage{amsmath}
\usepackage{graphicx}
\usepackage{xargs}[2008/03/08]
\usepackage[unicode=true,pdfusetitle,
 bookmarks=true,bookmarksnumbered=false,bookmarksopen=false,
 breaklinks=false,pdfborder={0 0 1},backref=section,colorlinks=false]
 {hyperref}
\hypersetup{
 draft,implicit=false}

\makeatletter

\usepackage{ol2}

\usepackage{xargs}[2008/03/08]

\usepackage{babel}

\makeatother

\begin{document}

\newcommandx\ket[1][usedefault, addprefix=\global, 1=]{|#1\rangle}
 \newcommandx\bra[1][usedefault, addprefix=\global, 1=]{\langle#1|}
 \newcommandx\avg[1][usedefault, addprefix=\global, 1=]{\langle#1\rangle}
 \newcommandx\var[1][usedefault, addprefix=\global, 1=]{\langle(\Delta#1)^{2}\rangle}

\twocolumn[ 

\title{High-fidelity spatially resolved multiphoton counting \\
for quantum imaging applications }

\author{Rados\l{}aw Chrapkiewicz,$^{1,*}$ Wojciech Wasilewski,$^1$ and Konrad Banaszek$^{1}$}

\address{
$^1$Faculty of Physics, University of Warsaw, ul. Ho\.z{}a 69, Warsaw, Poland \\
$^*$Corresponding author: radekch@fuw.edu.pl
}

\begin{abstract}
We present a method for spatially resolved multiphoton counting based
on an intensified camera with the retrieval of multimode photon statistics
fully accounting for non-linearities in the detection process. The
scheme relies on one-time quantum tomographic calibration of the detector.
Faithful, high-fidelity reconstruction of single- and two-mode statistics of multiphoton
states is demonstrated for coherent states and their statistical mixtures.
The results consistently exhibit classical values of Mandel and Fano
parameters in contrast to raw statistics of camera photo-events.
Detector operation is reliable for illumination levels up
to the average of one photon per an event area, substantially higher
than in previous approaches to characterize quantum statistical properties
of light with spatial resolution.
\end{abstract}


] 

\date{\today}

\maketitle
\noindent Spatial degree of freedom of light is a vital and applicable
resource to realize many quantum protocols \cite{Walborn2006,Tasca2011}.
In parallel, photon number resolved detection becomes increasingly
important in quantum protocols \cite{Hadfield2009,Eisaman2011,YuSpasibko2014}. In
both cases dedicated diagnostic techniques need to be developed. So
far, multiphoton states of light have been investigated with the use
of photon number resolving (PNR) detectors such as loop detectors
\cite{Rehacek2003,Achilles2004a}, multipixel photon counters \cite{Afek2009}
or transition-edge sensors \cite{Gerrits2012}. In parallel recent
advances in camera systems enabled experiments in entanglement imaging
\cite{Edgar2012,Fickler2013}, ghost imaging \cite{Aspden2013} and
sub-shot noise imaging \cite{Brida2010}. Single photon sensitive
cameras such intensified CCD (ICCD) and electron multiplied CCD (EM
CCD) have been also utilized as PNR detectors however their use was
severely limited to low illumination levels imposing trade-off between
high-spatial resolution or PNR capability \cite{Haderka2005,Blanchet2008,Edgar2012,Tasca2013,Machulka2014}. 

In this Letter, we introduce and demonstrate experimentally a versatile
method to measure quantum statistical properties of optical radiation
with spatial and photon number resolution. The method utilizes a single
photon sensitive camera whose pixels are grouped at the post-processing
stage into tiles -- ``macropixels'' that provide photon number resolving
capability. Individual tiles are calibrated using techniques developed
for quantum detector tomography \cite{Lundeen2008,Feito2009} which
fully take into account the non-linear response of the camera, including
saturation effects. Our approach extends the operation of the detector
to substantially higher illumination levels than could be handled
in previous experiments \cite{Haderka2005,Blanchet2008,Edgar2012,Tasca2013,Machulka2014}.
In most of these works, the raw statistics of camera photo-events
was directly identified with the input photon number distribution.
Such an assumption misinterprets cases when two or more photons generate
a single photo-event, leading to artificial non-classical effects
in photon statistics \cite{Bartley2013} when the detector
is illuminated with classical light. Moreover, joint measurements at multiple spatial locations
can yield sub-Poissonian correlations for such classical input states.
As a result, unambiguous verification of non-classical features of
detected radiation becomes difficult if not impossible. Our method
avoids these issues altogether by tomographic calibration of the detection process in a single experimental run 
and appropriate raw data processing. Characteristics of the detector such as the effective spatial resolution and the accessible photon number range can be defined at the postprocessing stage. 

\begin{figure}[b]
\includegraphics[width=1\columnwidth]{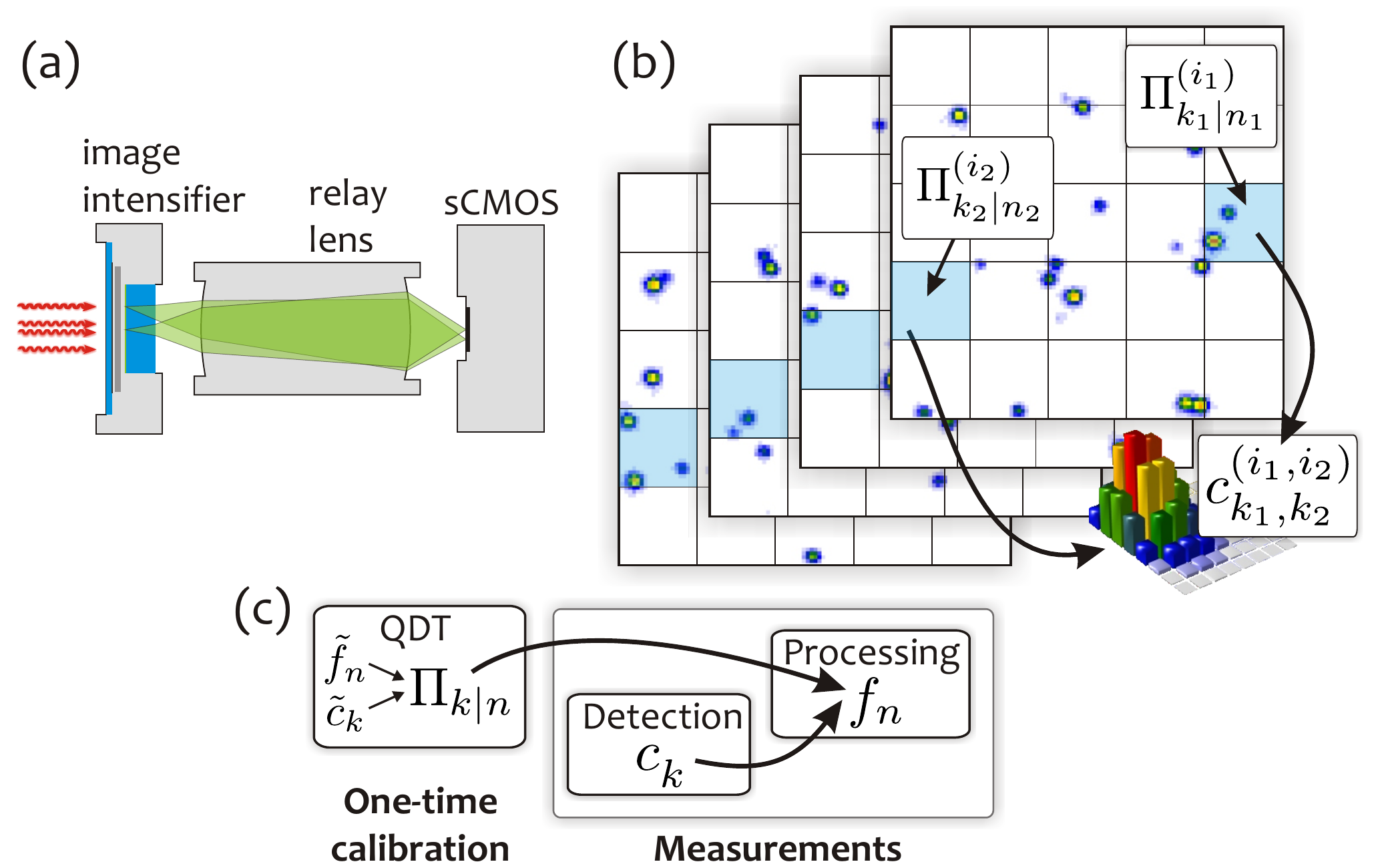}\centering

\caption{(a) Schematic of the detection on intensified sCMOS camera. (b) Camera
detection area with photo-events bright spots is arbitrarily divided
on tiles array in postprocessing stage. (c) Operation procedure initialized
by one-time tomographic calibration yields conditional probabilities
$\Pi_{k|n}$ for each tile. $\Pi_{k|n}$ allow to reconstruct initial
photon number distributions $f_{n}$ from raw photo-events statistics
$c_{k}$ in subsequent measurement runs. \label{fig:setup_idea} }
\end{figure}

\begin{figure}[t]
\includegraphics[width=1\columnwidth]{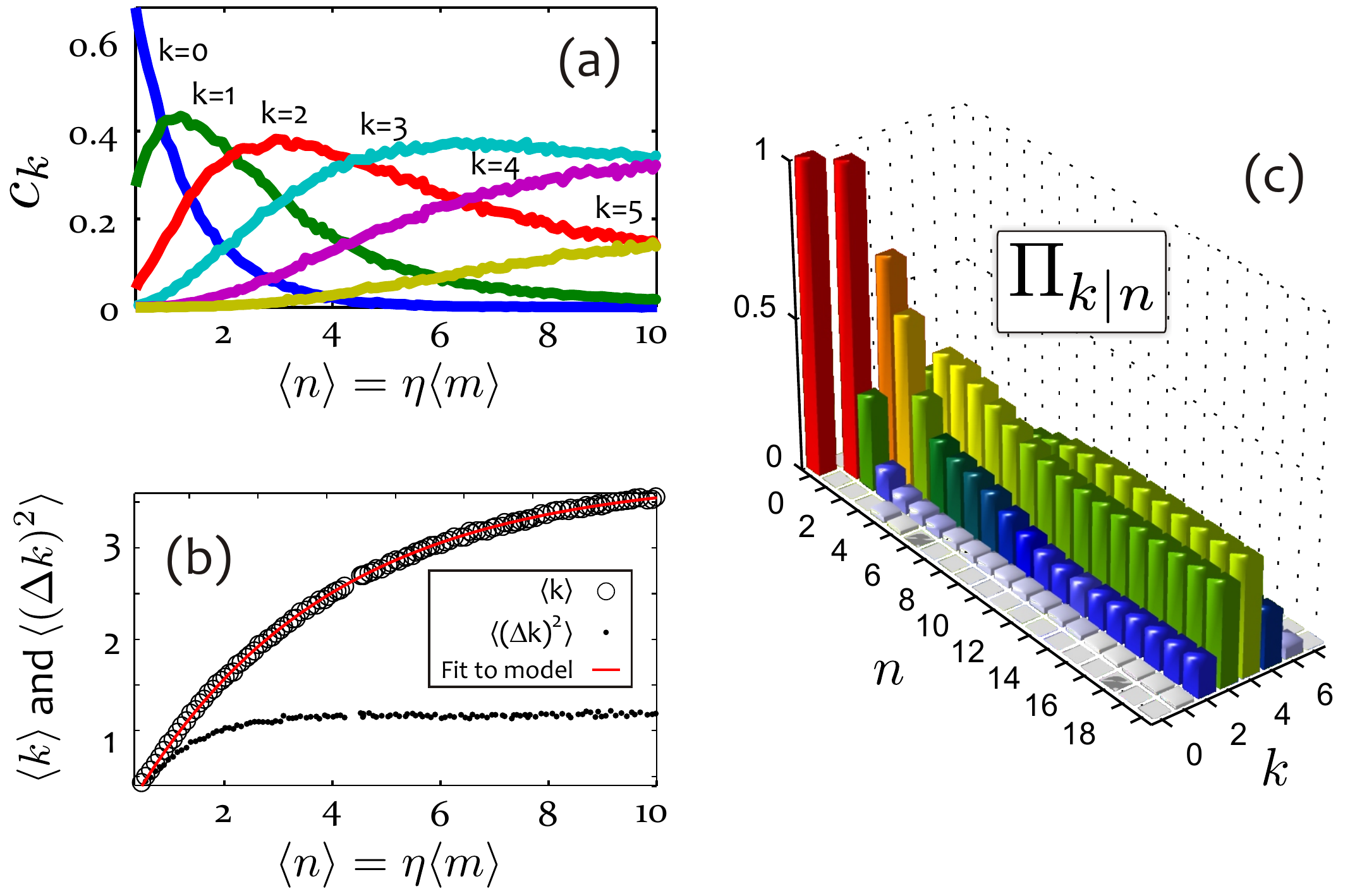}\centering
\caption{Exemplary operation of a single tile. (a) Photo-events statistics
$c_{k}$ measured with coherent state illumination versus mean number
of photons $\avg[n]=\eta\avg[m]$.
(b) Mean $\avg[k]$ and variance $\var[k]$ of photo-events.
Fit to the model Eq. (\ref{eq:model}) yielded $N=3.8$. (c) Quantum
detector tomography \cite{Lundeen2008,Feito2009} resulted in conditional
probability values $\Pi_{k|n}$ fully determining the tile operation. \label{fig:StatystykiZliczenPOVM}}
\end{figure}

Our experimental setup, depicted in Fig.~\ref{fig:setup_idea}(a)
is based on an intensified sCMOS (I-sCMOS) camera which operates similarly
to conventional ICCD cameras. Incident photons induce photoelectron
emission at the photocathode of the image intensifier. Each generated
photoelectron initiates a macroscopic avalanche in a micro-channel
plate, subsequently converted into an intense localized flash on the
phosphor screen. Flashes are registered as 5-pixel full width at half
maximum spots on a sCMOS sensor. A typical image on the sensor is
shown Fig.~\ref{fig:setup_idea}(b). The average intensity of the
brightest pixel is approx.\ 500 times higher than the camera noise
level. A spot is identified in real time by software as an individual
photo-event via its central pixel which has to be brighter than all
its neighbors up to a three pixel radius, and to exceed the noise
level at least five times. The precise location of the spot is found
by fitting a two dimensional paraboloid to the logarithm of the pixel
intensity within the same three pixel radius.

The spatial resolution of the detection system is defined at the postprocessing
stage by dividing the acquisition region into tiles, as exemplified
in Fig.~\ref{fig:setup_idea}(b). Each tile, labeled with an index
$i$, will be treated as a PNR quantum detector described by a conditional
probability distribution $\Pi_{k|n}^{(i)}$ which relates the number
$k$ of registered photo-events to the number $n$ of incident photons
that generated photoelectrons. For a single tile, the photo-event
statistics $c_{k}^{(i)}$ depends on the photon number distribution
$f_{n}^{(i)}$ through a standard expression 
\begin{equation}
c_{k}^{(i)}=\sum_{n}\Pi_{k|n}^{(i)}f_{n}^{(i)}.\label{eq:stat1D}
\end{equation}
This formula can be readily generalized to multiple tiles, which e.g.\ for
a pair of tiles $(i_{1},i_{2})$ takes the form 
\begin{equation}
c_{k_{1},k_{2}}^{(i_{1},i_{2})}=\sum_{n_{1},n_{2}}\Pi_{k_{1}|n_{1}}^{(i_{1})}\Pi_{k_{2}|n_{2}}^{(i_{2})}f_{n_{1},n_{2}}^{(i_{1},i_{2})}.\label{eq:stat2D}
\end{equation}
where $c_{k_{1},k_{2}}^{(i_{1},i_{2})}$ and $f_{n_{1},n_{2}}^{(i_{1},i_{2})}$
are respectively the joint photo-event statistics and the joint photon
number distribution. If the input state of light is known, the relation
given by Eq.~(\ref{eq:stat1D}) can be applied to find the conditional
probabilities $\Pi_{k|n}^{(i)}$ using the general methods of quantum
detector tomography (QDT) \cite{Lundeen2008,Feito2009}. Resorting
to QDT-based calibration fully takes into account the specifics of
the camera operation and extends the ability of the setup to resolve
higher photon numbers. Once the coefficients $\Pi_{k|n}^{(i)}$ are
found in a one-time procedure, Eqs.~(\ref{eq:stat1D}), (\ref{eq:stat2D})
and their multidimensional generalizations can be used to reconstruct
spatially resolved photon number distributions from joint photo-event
statistics. The operation procedure is summarized in Fig. \ref{fig:setup_idea}(c). 

In order to perform QDT we used as probe states 100~ns--10~$\mu$s
laser pulses excised from an acousto--optical modulator from a continuous
wave laser beam operating at a wavelength 795 nm. The quantum efficiency
of the photocathode at this wavelength was measured to be $\eta=20\%$.
The laser beam was attenuated by a set of calibrated neutral density
(ND) filters and its power was monitored on calibrated photodiode
along all measurements. The beam was shaped into circular flat-top
profile uniformly illuminating the acquisition region. Postselected
parts of the region served as model tiles. For an individual tile
we measured the photo-event statistics $c_{k}$, its mean $\avg[k]$
and variance $\var[k]$ for different mean number of incoming photons
$\avg[m]$ as presented in Fig. \ref{fig:StatystykiZliczenPOVM} (a-b).
The plots are parametrized with the number of incoming photons $\avg[m]$
related to the number of generated photoelectrons $\avg[n]=\eta\avg[m]$.
Each data point was obtained from acquisition of 10$^{4}$ frames.

We found that the mean number of photo-events $\avg[k]$ fits to the
model of $N$ on-off detectors with $\mathrm{QE}=\eta$ uniformly
illuminated by coherent state \cite{Sperling2012}:

\begin{equation}
\avg[k]=N(1-e^{-\eta\avg[m]/N}).\label{eq:model}
\end{equation}

Here the fit parameter $N$ plays a role of the equivalent number
of on-off detectors related to the single photo-event spots areas.
$N$ is proportional to the tile size which in our system is approximately
$N\times25$ sCMOS pixels. Because the absolute measurement of a mean
photon number $\avg[m]$ using calibrated ND filters and the photodiode
could be realized only for the entire laser beam, we repeated fits
to Eq. \eqref{eq:model} to determine precisely the fraction of light
incident of a specific tile.

\begin{figure}[t]
\includegraphics[width=1\columnwidth]{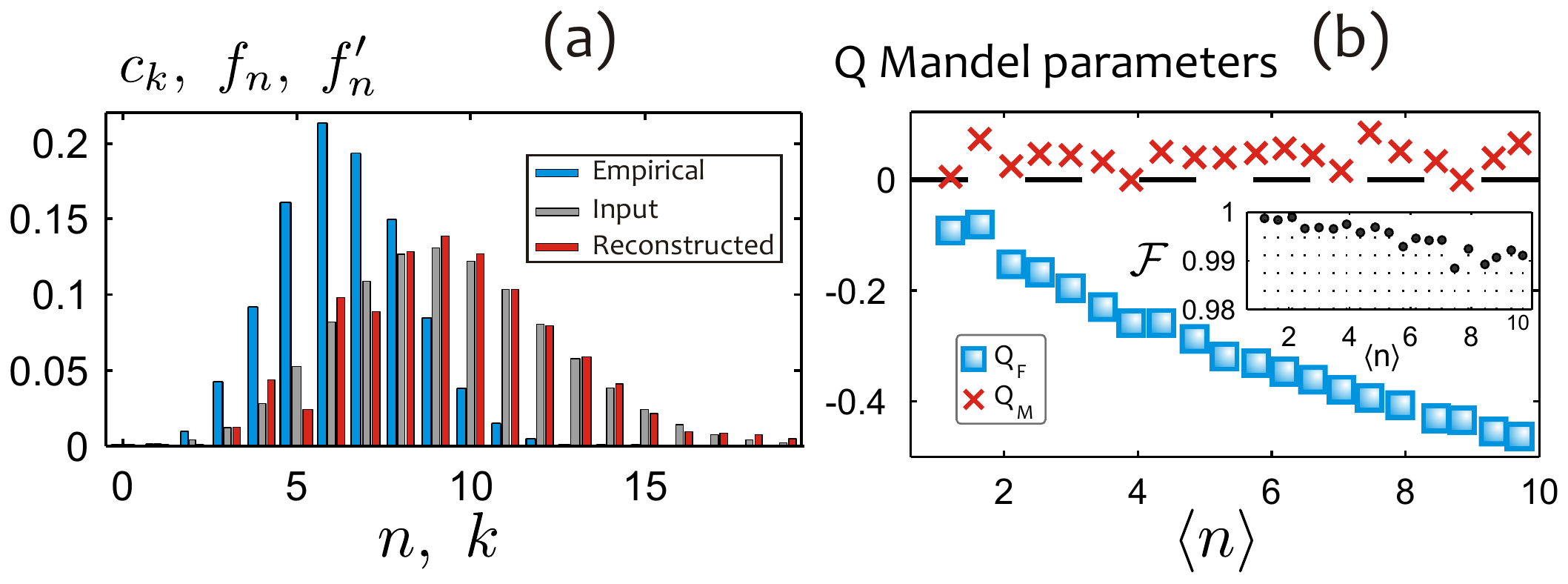}\centering

\caption{(a) Exemplary reconstruction of photo-event statistics $f'_{n}$ based
on measured counts statistics $c_{k}$ at tile of size $N=12$. Mean number of photons $\avg[n]=9.3$,
mean number of photo-events $\avg[k]=6.5$. (b) Mandel parameters
for photo-events $Q_{F}$, for reconstructed statistics $Q_{M}$ for
coherent states with variable mean photon number $\avg[n]$.
In the box fidelity of reconstruction.\label{fig:Rekonstrukcja1D}}
\end{figure}

The measured photo-event statistics $c_{k}$ (Fig. \ref{fig:StatystykiZliczenPOVM}(b))
and the Poissonian statistics $f_{n}$ of the probe states were used
to find the conditional probabilities $\Pi_{k|n}$ of the single tile
in QDT by numerical convex optimization algorithm \cite{Lundeen2008,Feito2009}.
The result of QDT is depicted in Fig. \ref{fig:StatystykiZliczenPOVM}
(c). It is seen that a tile of the intensified camera saturates in
a different manner than typical multiplexed detector. For large $n'>n_{\mathrm{sat}}$
the tile returns a stochastic number of photo-events $k$ given by
probability distribution $\Pi_{k|n'}=\Pi_{k|n_{\mathrm{sat}}}$. This
also implies that the variance of the number of photo-event counts
$\var[k]$ converges to the nonzero value as we observed in Fig. \ref{fig:StatystykiZliczenPOVM}(a). 

We used the conditional probabilities $\Pi_{k|n}$ to reconstruct
statistics of the known, test states to validate and quantify the
effectiveness of our method. The reconstruction process relies on
convex optimization algorithms analogously to QDT \cite{Lundeen2008,Feito2009}
with $f_{n}$ as the unknowns to be determined. As test states we
employed coherent states available at hand and their statistical mixtures. 

For one mode statistics we evaluate Mandel parameter $Q_{M}=\avg[(\Delta n)^{2}]/\avg[n]-1$
while for two mode, joint statistics we evaluate the Fano noise reduction
factor: $R=\avg[(\Delta(n_{1}-n_{2}))^{2}]/(\avg[n_{1}]+\avg[n_{2}]).$
For classical states we always have $Q_{M}\geq0$ and $R\geq1$. Because
the test states lay on the boundary between classical and non-clasical
states, they provide a sensitive probe into saturation effects. As
a qualitative measure of the effectiveness of our method, we evaluated
the fidelity $\mathcal{F}$ ~between the input and the reconstructed
statistics.

At first we performed state reconstruction for one mode coherent state
illuminating the tile of the size determined by equivalent number
of on-off detectors equals $N=12$ found from Eq. \eqref{eq:model}.
In Fig. \ref{fig:Rekonstrukcja1D} (a) we compare measured empirical
photo-event distribution $c_{k}$ with \emph{a priori }known input
statistics $f_{n}$ and reconstructed statistics $f_{n}'.$ Here the
Mandel parameter evaluated directly for the measured photo-event statistics
$Q_{F}=\avg[(\Delta k)^{2}]/\avg[k]-1=-0.45$. After reconstruction
$Q_{M}=0.04$ is classical and close to the expected, zero value.
In Fig. \ref{fig:Rekonstrukcja1D} (b) we present Mandel $Q$ parameters
and fidelity of reconstruction $\mathcal{F}>0.99$ for test states
of different mean values $\avg[n]$. Note that $Q_{F}$ is always
negative, even for illumination level of 0.05 photons per single photon
detector area. 

\begin{figure}[b]
\includegraphics[width=6.5cm]{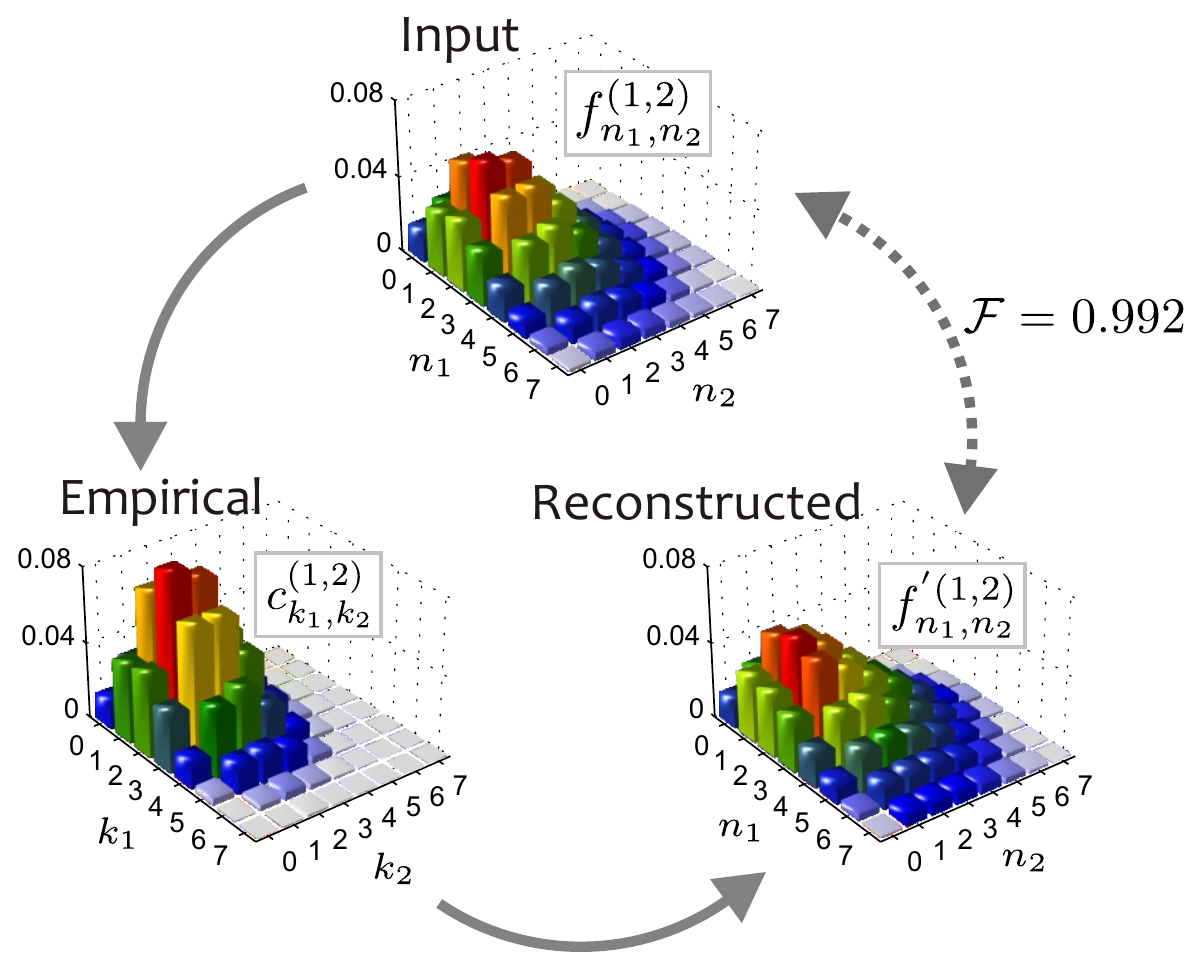}\centering
\caption{Histograms of joint statistics related to pair of tiles. The input state $f_{n_{1},n_{2}}^{(1,2)}$,
raw measured data $c_{k_{1},k_{2}}^{(1,2)}$ and reconstructed state
$f_{n_{1},n_{2}}^{'(1,2)}$ with high fidelity $\mathcal{F}$. Two calibrated tiles with $N^{(1)}=4.9$ and
$N^{(2)}=6.3$ were illuminated by equiprobably alternated statistical mixture of coherent
states, with $\bar{n}_{1}=2$, $\bar{n}'_{1}=3.7$.  \label{fig:Rekonstrukcja2D}}
\end{figure}

\begin{figure}[t]
\includegraphics[width=1\columnwidth]{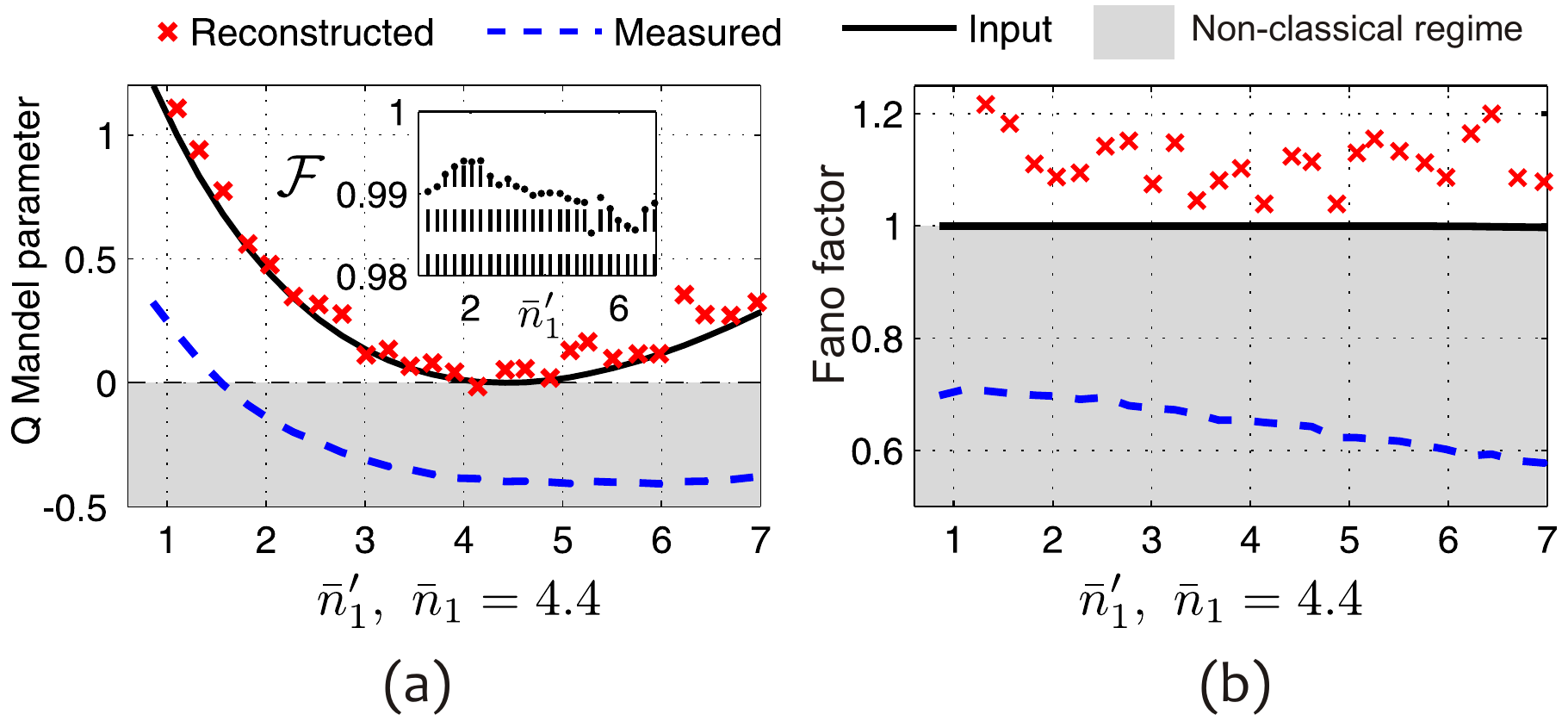}\centering

\caption{Results of joints statistics reconstruction of equiprobably alternated
statistical mixtures of coherent states.  (a)
Mandel parameters for photo-events $Q_{F}$ are superpoissonian for $\bar{n}'_{1}<1.5$. (b) Noise reduction Fano factor
for photo-events is always non-classical.  Reconstructed values consistently remain classical and are close to the expected values.\label{fig:MiaryNieklasycznosci2D}}
\end{figure}

Further insights into the operation of our method can be gained from
the reconstruction of joint statistics, where artificial sub-shot-noise
correlation can easily occur without proper reconstruction procedures.
As a test state we used equiprobable switched pairs of coherent states
simultaneously illuminating two tiles, labeled $i=1,2$. The mean
number of photons $\avg[n_{i}]$ at the $i-$th tile was switched
between $\bar{n}_{i}$ and $\bar{n}'_{i}$. The resulting state has
non-trivial joint statistics exhibiting classical correlations of
photon numbers between subsystems. Its Fano noise reduction factor
is $R=1$ and Mandel parameters of marginal statistics are $Q_{M}>0$
for $\bar{n}_{i}\neq\bar{n}'_{i}$. We postselected two similar but
not identical tiles which were calibrated together in QDT process
yielding the equivalent number of on-off detectors $N^{(1)}=4.9$
and $N^{(2)}=6.3$ and conditional probabilities $\Pi_{k_{1}|n_{1}}^{(1)}$
and $\Pi_{k_{2}|n_{2}}^{(2)}$ respectively. We checked there was
no crosstalk between the tiles. We present a typical reconstruction
result in Fig. \ref{fig:Rekonstrukcja2D} with high fidelity $\mathcal{F}=0.992$.
Note we succeeded in fixing artificial sub-poissonian correlation measured in photo-event
count from $R=0.71$ to $R=1.01$ value after reconstruction. In Fig.
\ref{fig:MiaryNieklasycznosci2D} we plot measured and the reconstructed
Mandel parameters $Q_{M}$ and $Q_{F}$ and the Fano noise reduction
factor for a variety of test states of the above type for fixed number
of photoelectons $\bar{n}_{1}=4.4$ as a function of $\bar{n}'_{1}$
. Note for each test state the Fano factor for photo-events was always
artificially non-classical and fixed by the reconstruction.

For $\bar{n}'_{1}<1.5$ we observe the particularly interesting situation
in which the directly measured marginal statistics are superpoissonian
($Q_{F}>0$ ) and the joint statistic is subpoissonian ($R<1$). Note
that this mimics the two mode squeezed state or their mixture. In
both cases the marginal statistics are superpoissonian thermal or
multimode thermal. Without the reconstruction procedure the correlated
classical state which we used is easily mistaken with its quantum
counterpart.

Both the single mode statistics and the two mode joint statistics
were reliably reconstructed up to the light intensity level corresponding
to one photon per single photon detection area $\avg[n_i]=(\bar{n}{}_{i}+\bar{n}'{}_{i})/2\simeq N^{(i)}$.
In this intensity regime we were able to remove artificial non-classical
effects from the reconstructed statistics and the fidelity of reconstruction
consistently exceeded $\mathcal{F}>0.99$.

These observations set the limit of useful operation of our method
up to one photon per single photo-event detection area on average.
It is two orders of magnitude higher than recently reported values
\cite{Edgar2012,Tasca2013}. This enhancement is enabled by proper
one-time quantum detector tomography and proper numerical state reconstruction.
The higher permissible light intensity means increased maximum photon
number resolving capability or, by dense tiles subdivision, increased
spatial resolution of the imaging system. For instance our particular
system can be used as an array of 10000 tiles capable of detecting
up to 6 photons on average, each at a frame rate of 200 Hz. The dark
count rate equals $6\cdot10^{-6}$ per each 100 ns gate duration per
tile and can be further reduced by two orders of magnitude by cooling
a photocathode.

In conclusion, we devised and demonstrated a novel method for measuring
multiphoton statistics with spatial resolution using intensified sCMOS
detector. Our method allowed in the high-illumination regime to retrieve
input photon statistics with high fidelity $\mathcal{F}>0.99$, free
from corruptible saturation effects. We verified the method using
one and two mode classical states consistently yielding classical
values of Mandel and Fano parameters for one and two dimensional statistics
respectively. The method consists in one-time calibration with appropriate
quantum detector tomography (QDT) yielding conditional probabilities
determining the operation of the camera tiles. Data processing in
our method is universal owing to application of QDT techniques, avoiding
altogether the need to develop analytical hardware description \cite{Haderka2005}
which is apparatus-specific and may often become overly complex. We
established the applicable limit of operation of the method to the
average illumination level of one photon per single photon detection
area far beyond the current limitations \cite{Edgar2012,Tasca2013}.
Our results pave the way to efficient and faithful imaging of multiphoton
quantum states of light.

We acknowledge R. Demkowicz-Dobrza\'{n}ski, C. Radzewicz and J. Iwaszkiewicz.
Project was financed by the National Science Centre no. DEC-2013/09/N/ST2/02229,
DEC-2011/03/D/ST2/01941, Polish NCBiR under the ERA-NET CHIST-ERA
project QUASAR.


\end{document}